\documentclass[referee, rnote]{aa}

\usepackage{txfonts}
\usepackage[ansinew]{inputenc}
\usepackage{textcomp}
\usepackage{graphicx}
\usepackage{latexsym}

\begin{document}

\title{On the orientation of the Sagittarius A* system}

\author{L. Meyer\inst{1}\thanks{Fellow of the International Max Planck Research School (IMPRS) for Radio and Infrared Astronomy at the Universities of Bonn and Cologne.} \and R. Sch\"odel\inst{1} \and A. Eckart\inst{1,2} \and W. J. Duschl\inst{3,4} \and V.~Karas\inst{5} \and M. Dov\v{c}iak\inst{5}  }

\institute{I.Physikalisches Institut, Universit\"at zu K\"oln, Z\"ulpicher Str. 77, 50937 K\"oln, Germany \and Max-Planck-Institut f\"ur Radioastronomie, Auf dem H\"ugel 69, 53121 Bonn, Germany \and Institut f\"ur Theoretische Physik und Astrophysik, Universit\"at zu Kiel, 24098 Kiel, Germany \and Steward Observatory, The University of Arizona, 933 N. Cherry Ave. Tucson, AZ 85721, USA \and Astronomical Institute, Academy of Sciences, Bo\v{c}n\'{i} II, CZ-14131 Prague, Czech Republic }

\date{Received  / Accepted }

%%%%%
%
\abstract {The near-infrared emission from the black hole at the Galactic center (Sgr~A*) has unique properties. The most striking feature is a suggestive periodic sub-structure that has been observed in a couple of flares so far.}
{Using near-infrared polarimetric observations and modelling the quasi-periodicity in terms of an orbiting blob, we try to constrain the three dimensional orientation of the Sgr~A* system.}
{We report on so far unpublished polarimetric data from 2003. They support the observations of a roughly constant mean polarization angle of $\sim 60\degr \pm 20\degr$ from 2004 -- 2006. Prior investigations of the 2006 data are deepened. In particular, the blob model fits are evaluated such that constraints on the position angle of Sgr~A* can be derived.}
{Confidence contours in the position -- inclination angle plane are derived. On a $3\sigma$ level the position angle of the equatorial plane normal is in the range $\sim 60\degr - 108\degr$ (east of north) in combination with a large inclination angle. This agrees well with recent independent work in which radio spectral/morphological properties of Sgr~A* and X-ray observations, respectively, have been used. However, the quality of the presently available data and the uncertainties in our model bring some ambiguity to our conclusions.}
{}


\titlerunning{On the orientation of the Sgr A* system}

\maketitle

\section{Introduction}

The near-infrared (NIR) regime plays an outstanding role in Galactic center research. The proper motion of stars visible in this waveband showed the existence of a supermassive black hole (BH) right in the center of our galaxy named Sagittarius~A* (Sgr~A*), see e.g. Eckart \& Genzel~\cite{eckigenzel}; Ghez et al.~\cite{ghez98}, \cite{gheznature}; Sch\"odel et al.~\cite{rainer1, rainer2}; Eisenhauer et al.~\cite{eisenhauer}. In 2003 also NIR emission directly from Sgr~A* has been detected (Genzel et al.~\cite{genzel}; Ghez et al.~\cite{ghez04}), which is the most underluminous BH accretion system so far accessible to observations (with a bolometric luminosity which is nine orders of magnitude lower than the Eddington luminosity). Short periods of increased radiation (called `flares') sometimes seem to be accompanied by a quasi-periodic oscillation (QPO), at least at $\lambda = 2.2\,\mu m$ (Genzel et al.~\cite{genzel}; Eckart et al.~\cite{ecki2}; Meyer et al.~\cite{ich2, ich}; see also Belanger et al.~\cite{belanger}; Aschenbach et al.~\cite{aschenbach1}; Yusef-Zadeh et al.~\cite{yusef}). However, note that all detections of a periodic sub-structure have been reported from observations with NACO/VLT only. While short timescale structure has also been reported from Keck observations (e.g. Ghez et al.~\cite{ghez05}), an independent confirmation of periodicity is still lacking (Ghez, priv. comm.). The suggestive QPO manifests itself as sub-flares with constant separation that are superimposed on a larger flare. Unfortunately, the exact cause of the QPO is far from clear. While the fact that the frequency of QPOs in BH binaries scale with one over the BH mass and that Sgr~A* seems to fit in this relation suggests Keplerian motion as the cause (Aschenbach~\cite{aschenbach2}, \cite{aschenbach3}; Abramowicz~\cite{abramowicz05}), recent magneto-hydrodynamic simulations disagree with that and instead identify pattern rotation as the source for the modulation (Chan et al.~\cite{chan}; Falanga et al.~\cite{falanga}). 

In this research note we focus on polarimetric measurements of Sgr~A* in the NIR (for polarization properties of Sgr~A* in the mm-regime see Marrone et al.~\cite{marrone2, marrone1}) and their interpretation in terms of the orbiting spot model, i.e. Keplerian motion in strong gravity is adopted as the cause for the sub-flares. Note that the orbiting blob model can be tested (and perhaps rejected), although the task cannot be achieved now, given the insufficient quality of data
available at present. In principle, constraints can be imposed on the
model by tracking all four Stokes parameters and comparing their time
evolution against the model. It is known that general relativity should
imprint specific features in the time evolution of a polarized signal
when a blob orbits near to a black hole (which is what we suggest here); the
direction of the polarization vector should wobble within a range
determined by the distance of the blob from the hole and the viewing angle
of the observer (e.g., Connors et al.~\cite{connors80}). Here, we discuss in particular the constraints that this modelling sets on the position angle of the normal to the equatorial plane of the spinning BH. 

In the next two sections we first report so far unpublished polarimetric data of Sgr~A* from 2003 that show that the mean polarization angle fluctuated only slightly for at least four years. Afterwards, this preferred direction is interpreted within the blob model.

\section{The data and their reduction}

The data we present here are polarimetric observations of Sgr~A* at $2.2\, \mu$m from October 2003\footnote{They are freely available on the ESO archive, program 072.B-0285(A)}. They have not been published before and are important to identify a favored orientation of the Sgr~A* system. They have been taken with the near-infrared camera and adaptive optics system NACO at ESO's Very Large Telescope (VLT) in combination with a wire-grid. The observations have been conducted in such a way that after $\sim 5$ min the wire-grid has been rotated. While it is now clear (Genzel et al.~\cite{genzel}; Eckart et al.~\cite{ecki2}) that this time resolution is too low due to the high variability of Sgr~A* in the NIR, nevertheless a mean polarization angle can be inferred. The data are of very high quality and show an exceptionally bright flare.

We carried out standard reduction techniques, i.e. sky subtraction, flat fielding and bad pixel correction. The point spread function was extracted on each individual image (Diolaiti et al.~\cite{diolaiti}) and then used for a Lucy-Richard deconvolution. After restoration with a Gaussian beam, aperture photometry on the diffraction limited images for individual sources with known flux and Sgr~A* was done.  For the extinction correction we assumed
$A_K=2.8$\,mag. Estimates of uncertainties were obtained from the
standard deviation of fluxes from nearby constant sources. The
calibration was performed using the overall interstellar polarization
of all sources in the field, which is 4\% at $25\degr$
(Eckart et al.~\cite{ecki95}; Ott et al.~\cite{ott}).

\begin{figure}[p]
\centering
\resizebox{10cm}{!}{\includegraphics[angle=270]{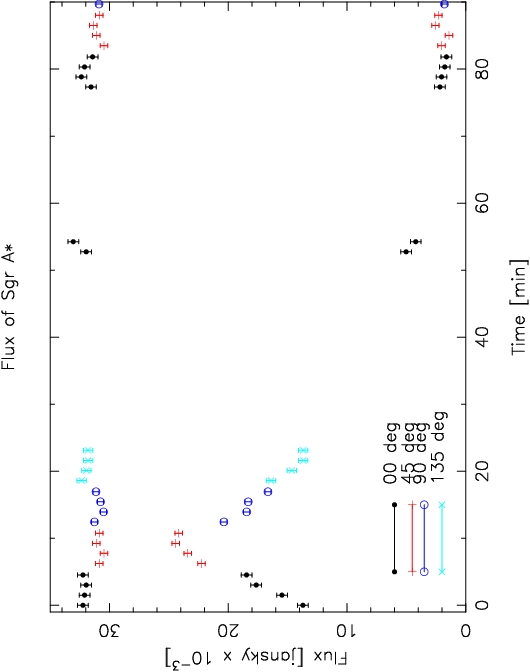}}
\caption{Dereddened flux of Sgr~A* and the constant reference star S2 (shifted 15\,mJy upward for better comparison). Note that the duration of the single sub-flare is $\sim 20$\,min. The flux for each channel is calibrated such that it matches the total flux, i.e. each Stokes parameter is individually calibrated to the total flux of the references sources and not the sum of two orthogonal angles. }
\label{flux}
\end{figure}

The dereddened flux of Sgr~A* and of a nearby constant star is shown in Figure~\ref{flux}. The flux was calibrated such that each angle seperately matched the total flux of known sources. That means that actually Figure~\ref{flux} shows approximately twice the flux for each angle. The first gap between $\sim 25-50$\,min is due to sky observations, the reason for the second gap is not traceable. The observations started exactly at the base of the peak.

\section{The mean polarization angle}

Figure~\ref{flux} shows the high variability of Sgr~A* with a very short rise and fall timescale consistent with previous observations. From these observations (Genzel et al.~\cite{genzel}; Eckart et al.~\cite{ecki2}; Meyer et al.~\cite{ich2}; Trippe et al.~\cite{trippe}) the following phenomenology of K-band flares has emerged: the first component is a broad underlying flare that lasts 50-120\,min. The second component is sub-flares that are superimposed on the broad flare and show a constant seperation of $17\pm 3$\,min. Having this context in mind and regarding the incompleteness of the data here, the single peak seen in the lightcurve in Figure~\ref{flux} may be interpreted as one sub-flare superimposed on an underlying flare. Note that although only one possible sub-flare can be seen, its duration is $\sim$20\,min and therefore exactly what is expected from previous observations that showed suggestive QPO activity.   

\begin{figure}[p]
\centering
\resizebox{10cm}{!}{\includegraphics[angle=0.3]{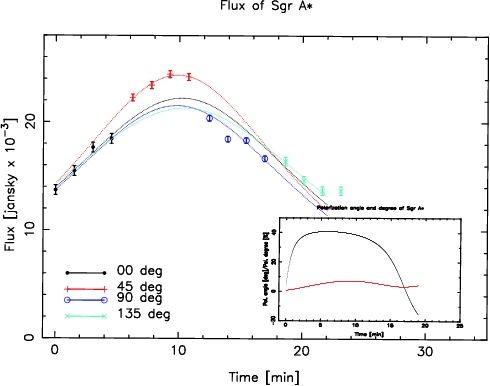}} 
\caption{Fits of a Lorentz profile to each channel of figure~\ref{flux}. With this approximate behavior, the polarization angle and degree of linear polarization can be inferred. They are shown in the insert in the lower right corner. The polarization angle is represented by the black line, the polarization degree is the red line.}
\label{fit}
\end{figure}

The polarimetric observing technique that was chosen for these observations here is certainly unsuitable as is known by now. The high variability demands the simultaneous measurement of all four position angles of the electric field vector. 
However, a shape of the sub-flare can be assumed and fitted to the data to allow a statement on the polarization angle and degree. Here, we approximate the sub-flare by a Lorentz profile of the form
\[
f(x)=\frac{s}{s^2+(x-t)^2}, \qquad s>0, \quad -\infty<t<\infty.
\] 
The choice of a Lorentzian to fit this part of the lightcurve is of course not unambigous. This clearly brings some uncertainty to the inferred polarization properties. Figure~\ref{fit} shows fits of a Lorentzian to each polarization angle. 

The polarimetric observations of Sgr~A* from 2004 -- 2006 (Eckart et al.~\cite{ecki2}; Meyer et al.~\cite{ich2}; Trippe et al.~\cite{trippe})  have shown that the degree of linear polarization and the polarization angle vary during a flare. The angle wobbles around a mean value during the high flux phase and then goes to different values in the decaying part of the flare (it is important to keep in mind that Sgr~A* is at $2.2 \mu m$ only detectable in its "flaring state" so that nothing is known about the polarization properties in its low flux phase).  The small insert in the lower right corner of Figure~\ref{fit} shows that the Lorentz profile fits reproduce this behavior from previous observations qualitatively. The mean angle can be read off to be $\sim 40\degr$ (east of north). 

Eckart et al.~(\cite{ecki2}) report a mean angle of $60\degr \pm 20\degr$ for observations in 2004 and 2005. While this is within the range inferred here, Meyer et al.~(\cite{ich2}) and Trippe et al.~(\cite{trippe}) arrive at a mean angle of $80\degr \pm 10 \degr$ and $80\degr \pm 25 \degr$, respectively, for the 2006 observing run. Concerning the uncertaintiy of the procedure adopted here, a conclusion that suggests itself is that the mean polarization angle of Sgr~A* changed only slightly during the past four years. This is especially true if one takes into account that a strict stability of the mean polarisation appears to be rather unphysical. Small fluctuations in the magnetic field configuration and/or a precession of the inner accretion disk seem likely. In this regard, our conclusion that the roughly constant mean polarisation points to a preferred position of the Sgr~A* system seems reasonable.

\section{The three dimensional orientation of the Sgr~A* system}

The existence of a favored polarization angle allows to investigate the orientation of the Sgr~A* system on the sky.
Here, we want to study it within the orbiting blob model. Meyer et al.~(\cite{ich2,ich}) calculated polarimetric lightcurves from an orbiting spot (here we use spot and blob interchangeably) around Sgr~A* and compared them to observations. This showed that this simple model, in which general relativistic effects on the radiation of a somehow confined, locally heated region plays the major role, leads to very good fits of the measurements. More precisely, in this model the sub-flares are due to a blob on a relativistic orbit
around the MBH, while an underlying ring accounts for the broad overall
flare. Relativistic effects like beaming, lensing, and change of
polarization angle imprint on the emitted intrinsic radiation
(e.g. Dovciak et al.~\cite{dovciak}; Connors \& Stark~\cite{connors}; Hollywood \& Melia~\cite{hollywood2}; Bromley et al.~\cite{bromley}; Falcke et al.~\cite{falcke}; Broderick \& Loeb~\cite{broderick1, broderick2}; Schnittman~\cite{schnittman}). In our model we assume that the variability
in the polarization angle and the polarization degree are only due to the
relativistic effects. As the emitted radiation of Sgr~A* is
synchrotron radiation (emitted in the disk corona), we assumed two different magnetic field
configurations to fit the light curves with our model. The first is
analogous to a sunspot and results in a constant $E$-vector
perpendicular to the disk. The second configuration is a global
azimuthal magnetic field that leads to a rotation of the $E$-vector
along the orbit. With this model at hand, observed polarimetric lightcurves can be fitted to investigate the parameters of the Sgr~A* system. The inclination angle, the dimensionless spin parameter $a_\star$, the brightness excess of the spot with respect to the disk, the initial phase of the spot on the orbit, the orientation of the equatorial plane on the sky, and the polarization degree of the disk and the spot are the free parameters.  In their discussion, Meyer et al.~(\cite{ich2,ich}) focused on the viewing angle and the spin parameter of Sgr~A*. In this note we extend the discussion to the position angle of the system, i.e. the angle of the equatorial plane normal.

\begin{figure}[p]
\centering
\resizebox{8cm}{!}{\includegraphics[angle=0]{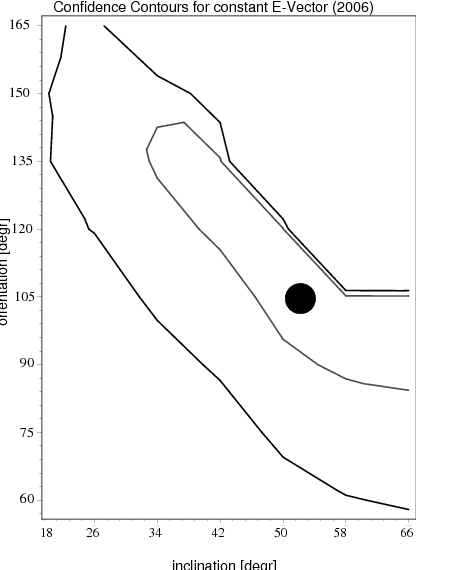}}
\caption{$1\sigma-$ and $3\sigma-$confidence contours calculated from fits on the 2006 data reported by Meyer et al.~\cite{ich2}. The vertical axis shows the position angle of the normal to the equatorial plane of the black hole on the sky (east of north). The horizontal axis shows the viewing angle (0\degr: face-on). The $\chi^2$-minimum is represented by the black dot. The assumed magnetic field configuration corresponds to the sun spot scenario (see text for details). }
\label{conf}
\end{figure}

Figure~\ref{conf} shows the $1\sigma-$ and $3\sigma-$confidence contours in the position angle ($\theta$) -- inclination angle ($i$) plane. The contours are results of the fits to the 2006 data presented in Meyer et al.~(\cite{ich2}). The magnetic field configuration corresponds to the sunspot case. The contours have been calculated such that on each point in the $\theta$ -- $i$ plane the $\chi^2$-minimum with respect to the other free parameters has been taken. The only exception is the dimensionless spin parameter $a_\star$ which has been fixed to $a_\star = 0.6$ throughout the calculations. The overall $\chi^2$-minimum lies then at $\theta = 105\degr$, $i=55\degr$ (marked with the black dot). As the contours show, this minimum is not unambigous, but it is nevertheless noteworthy that the formal $\chi^2$-minimum coincides exactly with the finding of Markoff et al.~(\cite{sera}). In their paper, they fitted spectral and morphological properties of Sgr~A* within the jet model. They also arrive at a position angle of 105\degr (east of north) for the jet axis together with an almost edge on viewing angle. The inclination angle in our work (Figure~\ref{conf}) may not be strict edge-on, but it has the trend to be. Also note that our approach can only cover the region $i\lessapprox 70\degr$. Furthermore, Muno et al.~(\cite{muno}) and Eckart et al.~(\cite{ecki2}) present X-ray and NIR data of one long, narrow feature of synchrotron-emitting particles that point toward Sgr~A* and may be identified as a jet. The angle of this jet-like feature on the sky is close to the $105\degr$ inferred here and thereby supports our finding.

It is interesting that different models and methods seem to converge to a viewing angle which is close to edge-on. Not only Markoff et al.~(\cite{sera}) and Meyer et al.~(\cite{ich2}) but also Falanga et al.~(\cite{falanga}) and Huang et al.~(\cite{huang}) find such a configuration. Using the Rossby wave instability model (Tagger \& Melia~\cite{tagger}), Falanga et al.~(\cite{falanga}) arrive at an inclination angle of $i \approx 77\degr$. Huang et al.~(\cite{huang}) noted that large inclination angles are preferred when a radiatively inefficient accretion flow model is coupled to a ray tracing algorithm and compared to recent VLBI size measurements. Taking the high inclination for granted, Figure~\ref{conf} shows that our model predicts a position angle of $\sim60\degr - 108\degr$. This quite broad range includes the result from Markoff et al.~(\cite{sera}) and Muno et al.~(\cite{muno}) as well as -- at least in the $3\sigma$ limit -- Muzic et al.~(\cite{kora}). The latter work investigated the proper motion of thin filaments at the Galactic center, which may be driven by some kind of collimated outflow. Sgr~A* is identified as one possible source for such an outflow and if this is true, a preferred ejection direction of $\sim 60\degr$ is possible.

\begin{figure}[p]
\centering
\resizebox{8cm}{!}{\includegraphics[angle=0]{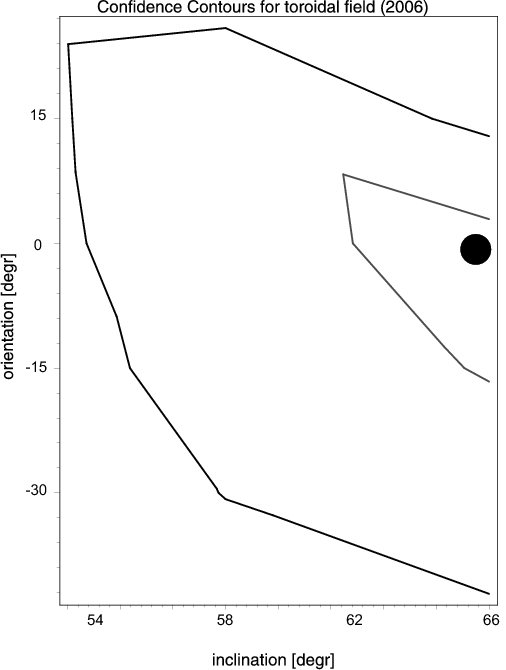}}
\caption{$1\sigma-$ and $3\sigma-$confidence contours calculated from fits on the 2006 data reported by Meyer et al.~\cite{ich2}. The vertical axis shows the position angle of the normal to the equatorial plane of the black hole on the sky (east of north). The horizontal axis shows the viewing angle (0\degr: face-on). The $\chi^2$-minimum is represented by the black dot. An azimuthal magnetic field has been assumed. }
\label{toro}
\end{figure}

It is only useful to show the fits to the 2006 data here. This is for two reasons. First, the confidence contours of the 2005 and 2006 data are very similar and second, the 2006 data are of far better quality than the 2005 data. But it is of major importance to note the shortcomings of our approach. First of all, the magnetic field geometry of the blob is completely unknown. As described above we used two simple approximations to describe the spot's field: (i) a sunspot like geometry in which the resulting $\vec{E}$-vector is always perpendicular to the equatorial plane, (ii) an azimuthal magnetic field, which is suggested by MHD simulations of magnetized accretion disks that often develop a toroidal
component of the magnetic field. The former leads to better fits, that is why we have discussed it at length. But the latter case can not be discarded. The corresponding confidence contours are shown in Figure~\ref{toro}. Unfortunately, its $\chi^2$-minimum lies at rather small position angles, so the above conclusions get weakened. However, this minimum lies $\Delta \chi^2 \approx 5$ higher than in the sunspot scenario ($\chi^2 \approx 3$ and $\chi^2\approx 8$, respectively; all values are reduced-$\chi^2$ values). Although this is not a $3\sigma$ exclusion, it makes the sunspot scenario more likely. Another caveat concerns the unknown foreground polarization. Its value at the position of Sgr~A* is important in the determination of the position angle. We had to assume that it is equal to the average value in the field. With these restrictions in mind, our work at least shows the consistency of previous independent findings with our model.

\section{Summary}

The purpose of this research note is to show that the mean polarization angle in the near-infrared was almost constant from 2003-2006, and to infer the position angle of Sgr~A* from this preferred direction in terms of the orbiting spot model. For a high inclination angle the $3\sigma$ range for the position angle is derived to be $\sim 60\degr - 108\degr$, thereby supporting the findings of Markoff et al.~(\cite{sera}), and -- less likely -- Muzic et al.~(\cite{kora}) who arrived at position angles of $\sim105\degr$ (see also Muno et al.~\cite{muno}) and $\sim60\degr$, respectively, with complementary methods. However, there are severe caveats. The unknown magnetic field structure, the uncertain foreground polarization, and the lack of a clear understanding of the detailed hydrodynamics of the blob challenge our conclusions.

\begin{acknowledgements}
We are grateful to Sera Markoff and Andrea Ghez for fruitful discussions and comments. This work was supported in part by the Deutsche Forschungsgemeinschaft (DFG) via grant SFB 494 and the Center for Theoretical Astrophysics in Prague. 
\end{acknowledgements}

\end{document}